# Investigation of spatiotemporal output beam profile instabilities from differentially pumped capillaries

**MARTIN GEBHARDT,**[1,2,*] **EMMANUEL B. AMUAH,**[1,3] **ROBERT KLAS,**[1,2] **HENNING STARK,**[1] **JOACHIM BULDT,**[1] **ALBRECHT STEINKOPFF,**[1] **AND JENS LIMPERT**[1,2,4]

[1]*Institute of Applied Physics, Abbe Center of Photonics, Friedrich-Schiller-Universität Jena, Albert-Einstein-Str. 15, 07745 Jena, Germany*
[2]*Helmholtz-Institute Jena, Fröbelstieg 3, 07743 Jena, Germany*
[3]*Current address: Department of Physics and Astronomy, Aarhus University, Ny Munkegade 120, DK-8000 Aarhus, Denmark*
[4]*Fraunhofer Institute for Applied Optics and Precision Engineering, Albert-Einstein-Str. 7, 07745 Jena, Germany*
*martin.gebhardt@uni-jena.de

**Abstract:** Differentially pumped capillaries, i.e., capillaries operated in a pressure gradient environment, are widely used for nonlinear pulse compression. In this work, we show that strong pressure gradients and high gas throughputs can cause spatiotemporal instabilities of the output beam profile. The instabilities occur with a sudden onset as the flow evolves from laminar to turbulent. Based on the experimental and numerical results, we derive guidelines to predict the onset of those instabilities and discuss possible applications in the context of nonlinear flow dynamics.

## 1. Introduction

The use of gas-filled, hollow-core waveguides has revolutionized fundamental research and applications in optics and photonics. Nowadays, they are an enabling component for nonlinear frequency conversion to the ultraviolet [1,2] or soft X-ray [3] spectral regions, they are used for enhanced gas spectroscopy [4] as well as for nonlinear pulse compression [5]. With the ever-growing performance of table-top, solid-state laser technology [6–10], the generation of extremely short and intense temporal waveforms through external pulse shortening is in high demand within the community [11]. Because of the broadband waveguiding capabilities, the straightforward experimental handling and the possibility to realize a broad range of core diameters on the order of 0.1 mm – 1 mm, glass capillaries have been widely used in the spectral broadening step of nonlinear pulse compression. This is especially true when the goal was to push the pulse durations of high-energy, femtosecond laser systems to the few-cycle regime [12–17]. In fact, it has been demonstrated recently that this approach is also scalable to hundreds of watts in average power [13,18]. Operating a hollow-core fiber compressor in a differential pumping scheme, i.e. implementing low pressure or vacuum at the optical input of the fiber and high pressure at its optical output, was established as a very helpful measure to achieve well-defined input coupling conditions without perturbations due to ionization or the spatial Kerr effect [12–14,16,17,19]. Furthermore, it can be shown that the gas particle density gradient along the fiber can be employed to wash out the phase-matching of unwanted parametric generation via position-dependent changes in the dispersion landscape. Another successful strategy to avoid detrimental ionization effects [20] associated with too high laser intensities is to use capillaries with very large core diameters (i.e. several hundred times the operation wavelength [16]). Clearly, this also leads to a low fundamental guidance loss for short capillary lengths and can be exploited to build very efficient fiber-based compressors [21]. However, at reasonable lengths of the straight fibers, which are certainly <10 m for standard laboratories, the gas throughput for such differentially pumped, large-core capillaries can become considerable at a certain pressure gradient. Because the strength of the nonlinear phase accumulation is closely related to the gas particle density, it is typically desired to increase the pressure at the optical output of the capillary as much as possible (ideally up to the self-focusing limit), while the region around the optical input is evacuated for the above described reasons. This scheme leads to the aforementioned strong pressure gradients and high throughputs. Interestingly, it has been reported that certain instabilities (of the optical output) occur, if the pressure gradient for a differentially pumped capillary with 350 µm core diameter and 1 m length exceeds 3 bar (nitrogen) [22], but the authors neither investigated this effect nor explained its origin. This raises the question of whether and to which extent a high gas throughput can have a detrimental impact on the output laser characteristics. Answering this question could be especially interesting for the wavelength (up-)scaling of nonlinear pulse compression using differentially pumped capillaries. This is because the higher possible gas pressures (due to the wavelength dependence of the self-focusing limit [23]) and the increased fiber dimensions (to allow for reasonable transmission and spectral broadening [20]) are not necessarily compatible with a low gas throughput.

In this work, we show that extensive gas volume flow through capillary waveguides can lead to dynamic changes of the output beam's spatial power distribution. In the experiments described herein, we found that the spatiotemporal fluctuations occur mostly in time scales in the 1 ms – 10 ms region. Furthermore, we have identified that the instability onset is closely linked to the Reynolds number, which is a benchmark for different volume flow conditions [24]. Based on the detailed observations in this work, including experiments with multiple geometries and gas species, we derive guidelines that will help to predict the onset of the observed instabilities, which are clearly harmful to laser applications. At the same time, this work opens up a potential route toward easy and straightforward experimental characterization of volume flow through capillaries. Such optical measurements will be a valuable tool for fundamental and applied research in the field of complex nonlinear flow dynamics [24–26].

## 2. Experimental setup and results

The principal outline of the experimental setup can be seen in Fig. 1(a). For the experiments presented herein, we used a mode-locked oscillator, that provided a femtosecond pulse train at a repetition rate of 40 MHz. The central wavelength of the pulse spectrum was around 1.03 µm. In the initial experiment, we launched 100 mW of average power to the capillary under test (500 µm inner diameter, 5 m length) and optimized the input coupling for maximum transmission and fundamental mode operation. In fact, the capillary dimensions were chosen such that they represent reasonable experimental conditions for high-energy nonlinear pulse compression. The transmitted power was 70 mW, which is about 80% of the expected power transmission of the waveguide's fundamental mode (the deviation from the theoretical transmission is attributed to coupling losses). In the experiments, the capillary was set up such that it connected two sealed vacuum/pressure chambers. The pressure in the chamber containing its optical input was held at $p_0$ = 10 mbar using a vacuum pump (XDS35i, Edwards, nominal pumping speed 35 m$^3$/h), while the upstream pressure $p_1$ at its optical output was varied.

We characterized the diverging output beam with a high-speed camera (HSC (V611, Phantom)) and with a fast photodiode (PD (PDA36A(-EC), Thorlabs)). For the photodiode measurements, only a fraction (∼20% of the total power) of the beam was selected with a stationary rectangular aperture, while most of the power was blocked (see Fig. 1(b)). This way, fluctuations in the mode profile translate to a temporally varying photodiode signal, detected with a fast oscilloscope (HDO6104, Teledyne Lecroy), from which comparable values such as a standard deviation or a frequency spectrum can be derived. Note that this procedure has been established to characterize the dynamics of transverse mode instabilities in high-power fiber amplifiers [27]. For the photodiode measurements presented herein, the maximum bandwidth limited by the electronic signal amplification was 1.6 MHz. As it will become clear in short, this is sufficient for the investigation of the temporal instabilities observed herein.

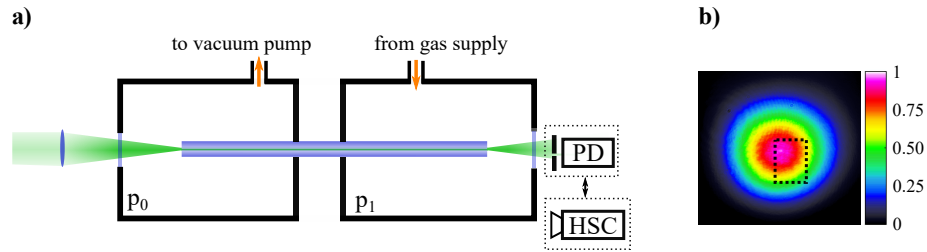

**Fig. 1.** a) Experimental setup consisting of the capillary under test, an evacuated, optical input vacuum chamber held at a pressure $p_0$ and an optical output pressure chamber held at a pressure $p_1$. The laser light was coupled to the fundamental mode of the differentially pumped capillary and the spatial output was characterized with a fast photodiode (PD) or a high-speed camera (HSC). b) Typical output beam profile and approximate position of the rectangular aperture (dashed line) employed in the photodiode measurements. The color bar shows the normalized counts.

In the initial experiment, nitrogen was used to create the pressure gradient and volume flow through the capillary under test. At first, the time domain photodiode signal was analyzed for an absolute pressure $p_1$ = 1.1 bar, i.e. a very small gas flow. At this point, we calculated a relative standard deviation of 0.2%, a value that is just above the detection limit of the measurement apparatus and mostly associated with the general intensity and beam pointing fluctuations of the laser source. Subsequently, the pressure $p_1$ at the optical output side was increased, leading to increasing gas throughput, while the photodiode signal was carefully monitored. Figure 2 presents the behavior of the standard deviation as the upstream pressure is increased. At an upstream pressure $p_1$ = 4.2 bar, the signal's standard deviation had increased dramatically, with a threshold-like onset.

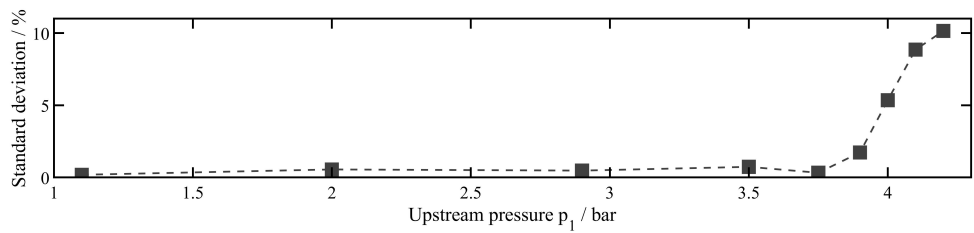

**Fig. 2.** Standard deviation of the photodiode signal for different upstream pressures. The data points are connected by the dashed line to guide the eye.

Subsequently, we have performed a more detailed evaluation of the measurements for $p_1$ = 4.2 bar and we compare them to a representative pressure value below the instability onset ($p_1$ = 2.9 bar). The photodiode signals for both cases are shown in Fig. 3(a) and 3(b). It is clear from the spectral domain analysis (Fig. 3(c) and 3(d)) of those photodiode traces, that significant noise was added at frequencies <10 kHz when the pressure $p_1$ was set to 4.2 bar. In this case, the integration of the power spectral density starting from the high frequency side (from 2 MHz down to 2 Hz) yields a relative intensity noise (RIN) of about 10% (Fig. 3(f)). In contrast to this, the spatial RIN for $p_1 = 2.9$ bar is only 0.4% (Fig. 3(e)), which is very close to the initially measured relative standard deviation of the temporal photodiode trace with negligible gas flow.

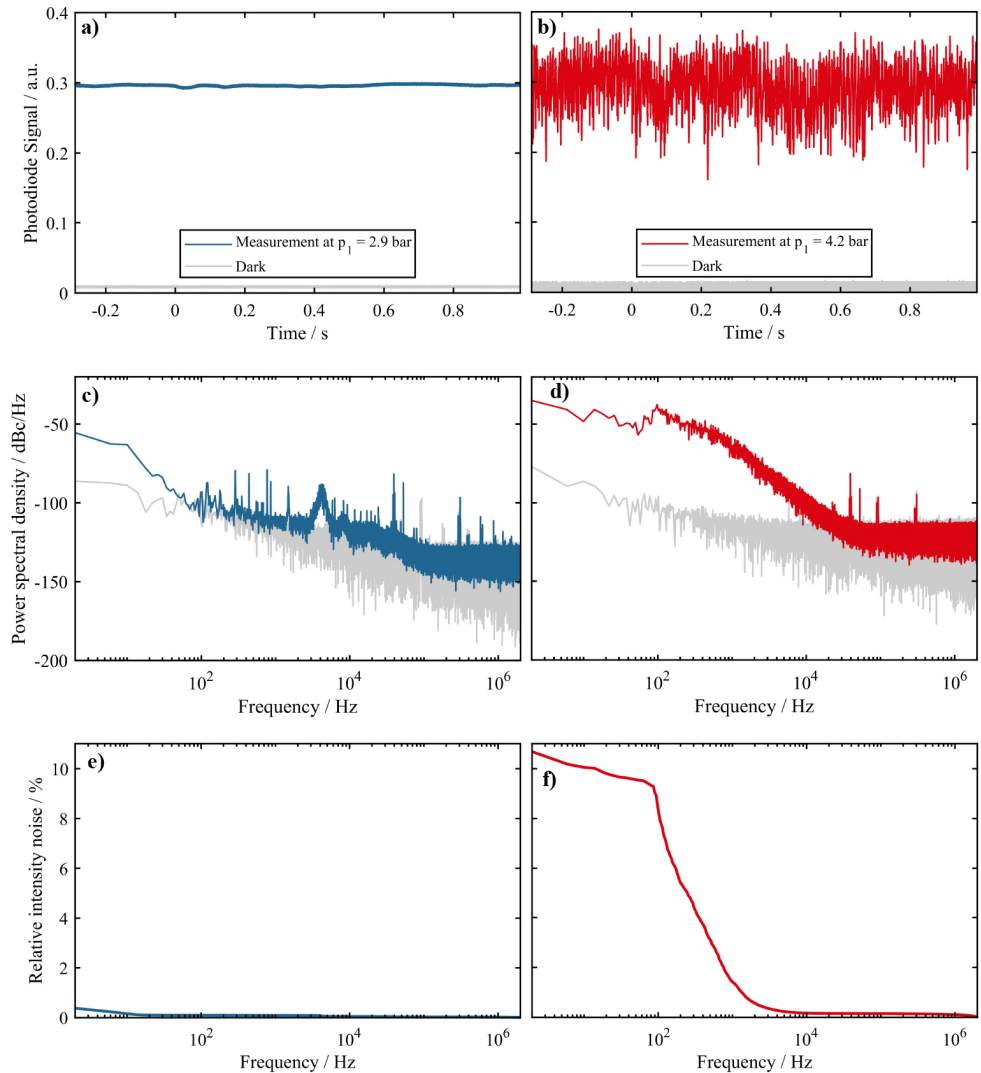

**Fig. 3.** Measured photodiode signals for a) $p_1 = 2.9$ bar and b) $p_1 = 4.2$ bar, together with c), d) corresponding power spectral densities. The difference in the dark traces results from the different scales in the analog-digital conversion, that are necessary for optimal resolution of the signal. e), f) Integrated relative intensity noise. Gas species: nitrogen.

It appears that most instabilities manifest themselves at frequencies between 0.1 kHz and 1 kHz. This enabled us to monitor the suspected spatiotemporal changes in the output beam profile using a high-speed camera with a 21 kHz frame rate and 5 µs integration time. Figure 4 presents an evaluation of the frames, captured at $p_1 = 2.8$ bar (below the instability onset) and $p_1 = 4.2$ bar (above the instability onset).

The evaluation of the high-speed camera data was done by calculating the position of the beam (the center of mass) in the horizontal (x-) and the vertical (y-) direction as well as the beam diameter (using the 4σ-method) in both of these axes. The comparison between Fig. 4(a) and 4(b) shows that above the threshold pressure, there is a significant temporal variation in the output spatial beam profile (e.g. compare frames 1824 and 3139, respectively). Generally speaking, it is found that much of the output power is still spatially confined in a central lobe, as could be expected from the fundamental mode of the capillary without strong pressure gradient. However, it can be observed that the beam area, center of mass and peak intensity undergo temporal fluctuations. Clearly, a laser output with such temporal and spatial properties is hardly usable, especially if the laser system operates at a repetition rate in the 0.1 kHz – 10 kHz range. In order to better understand this behavior and to potentially predict the instability threshold for different experimental conditions, it is important to investigate the origin of the above described observations.

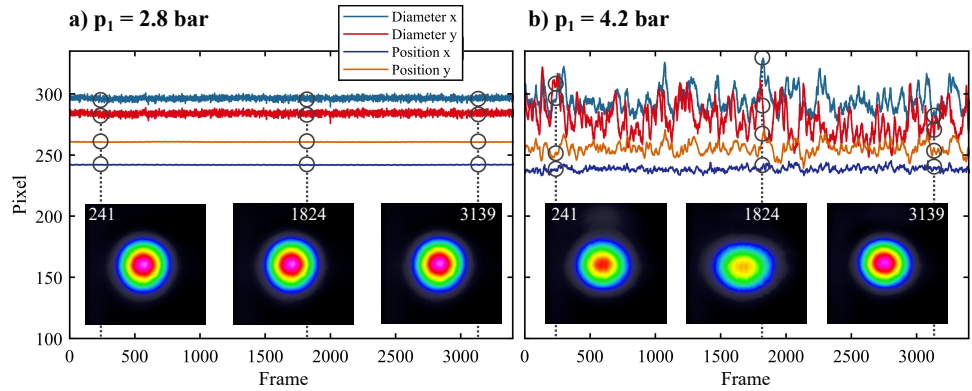

**Fig. 4.** Evaluation of high-speed camera frames (time between capturing two consecutive frames is 47.6 µs) for a) 2.8 bar at the optical output and b) 4.2 bar at the optical output. The variable x represents the horizontal direction and the variable y represents the vertical direction. Insets: representative beam profiles at frame numbers 241, 1824 and 3139 normalized to the maximum counts of frame 3139, respectively. The insets show the full 512 × 512 pixels of the recorded frames (pixel pitch: 20 µm). For the full high-speed videos, see Visualization 1 and Visualization 2, respectively. The real-time duration of the full high-speed videos is 0.164 seconds (2.8 bar at the optical output) and 0.389 seconds (4.2 bar at the optical output). The visualizations are rendered in slow-motion (350x).

## 3. Discussion

Because the observations described so far are clearly linked to the gas volume flow through the capillary, we have implemented an iterative numerical formalism to calculate the throughput and to estimate the Reynolds number *Re*. The theoretical basis for this is described in detail in Ref. [28]. Herein, we assume that the transition region between laminar and turbulent flow is located between $Re = 2040$ [24] and $Re = 3500$ [28]. Within these boundaries, the numerical method is especially important for the calculation of the throughput $Q$, while elsewhere, it is also possible to use the following explicit equations [26,29,30]:

$$Q_{\text{lam}} = \frac{\pi}{256} \cdot \frac{1}{\eta} \cdot \frac{d^4}{L} \cdot (p_1^2 - p_0^2), \qquad (1)$$

$$Q_{\text{tur}} = d \cdot \left(0.39\pi^2 \cdot \frac{d^3(p_1^2 - p_0^2)}{2L}\right)^{4/7} \cdot \left(\frac{RT}{M_{\text{mol}}}\right)^{3/7} \cdot \left(\frac{4}{\pi\eta}\right)^{1/7}. \qquad (2)$$

Here, $Q_{\text{lam}}$ is the throughput assuming laminar flow conditions, $\eta$ is the dynamic viscosity of the fluid, $d$ is the capillary inner diameter, $L$ is the capillary length. $Q_{\text{tur}}$ is the throughput assuming turbulent flow conditions, $R$ is the universal gas constant, $T$ is the temperature of the inflowing gas, $M_{\text{mol}}$ is its molar mass and the pressures $p_0$ and $p_1$ are defined above. In Ref. [28], the Reynolds number is defined as

$$Re = \frac{4 M_{\text{mol}} Q}{RTB\eta}, \qquad (3)$$

where $B = \pi d$ is the perimeter of the circular waveguide channel. With this formalism, it is possible to calculate the throughput and the Reynolds number for a differentially pumped capillary (nitrogen) with 5 m length and 500 μm inner diameter. The results of this calculation are shown in Fig. 5. We note that for $p_1 > 0.4$ bar, the downstream pressure is higher than $p_0 = 10$ mbar, because the flow is chocked.

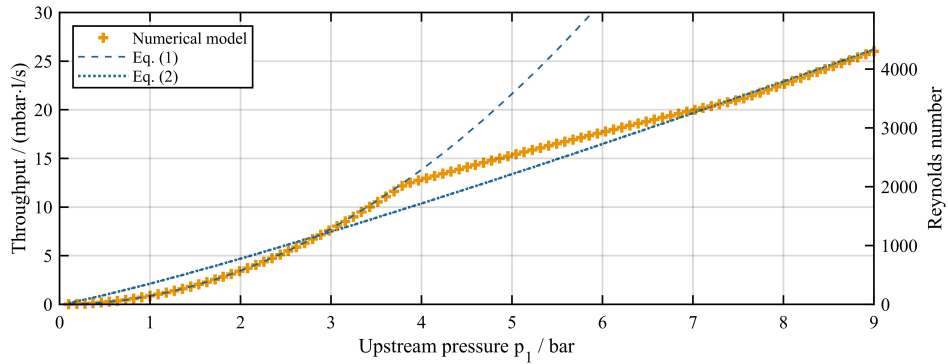

**Fig. 5.** Comparison of volume flow and Reynolds number based on analytic solutions for laminar flow (Eq. (1)), turbulent flow (Eq. (2)) and for the numerical model as described in Ref. [28] depending on the nitrogen pressure $p_1$. The capillary under consideration is 5 m long with 500 μm inner diameter.

It can be derived from the Reynolds number, that the above described measurements at around $p_1 \sim 3$ bar have been performed in the laminar flow regime ($Re<2000$). In contrast to this, we found that the upstream pressure $p_1 = 4.2$ bar, at which strong spatiotemporal instabilities of the output beam were observed, can be associated with the lower bound of the transition between laminar and turbulent gas volume flow through the capillary. This observation could be interpreted as follows. In the transition region, small disturbances can cause localized chaotic fluctuations in the flow, called puffs [31]. Despite their transient nature [24], we believe that those turbulences in the flow have a non-negligible impact on the local refractive index of the gas (via changes in the local gas particle density), which would affect the light propagating through the capillary. Effectively, the initially excited beam profile within the waveguide propagates through a series of non-uniform cross-sectional refractive index profiles, which will certainly change its appearance as compared to the unperturbed case. Clearly, this interpretation is to be confirmed by further studies and detailed theoretical work. Our focus is purely observational, and the key point is that we are able to present detailed evidence for the connection between the beam profile fluctuations and the Reynolds number. Subsequently, we have performed multiple measurements of the spatial RIN (as described in section 2) in the frequency interval 2 Hz – 2 MHz for different capillary geometries and gas species. The capillaries used in this work originate from three different manufacturers (as will become clear in short, all showed the same behavior). Their outer diameters are in the range of 1 mm to 3.5 mm. The capillary mounting was realized using either a commercially available stretched hollow-core fiber setup or a home-built setup for rigid, rod-type capillaries, in which the fibers could be fixed close to the input and/or output facets. The measurements are compared to the respective Reynolds numbers (calculated using the above-mentioned numerical methods) in Fig. 6.

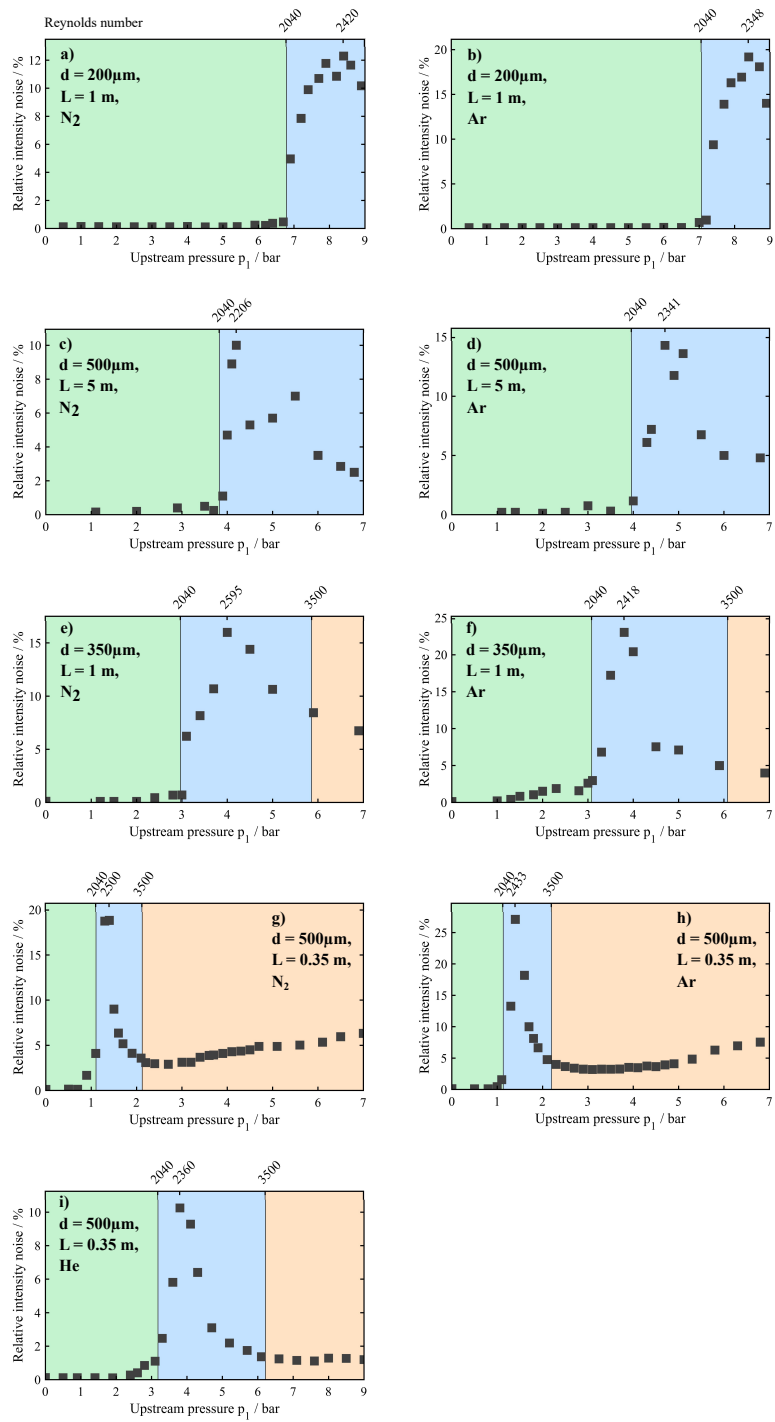

**Fig. 6.** Compilation of measured RIN values for 9 different experimental configurations as specified in a) to i). The length to diameter ratio ranges from 700 (g-i) to 10000 (c, d) with gas species argon, nitrogen and helium. The top x-axis represents the Reynolds number boundaries for the transition region between laminar and turbulent flow as well as the Reynolds number for maximum RIN. The colored background is to guide the eye with regard to the flow regimes.

The RIN measurements presented in Fig. 6 show, without exception, a significant increase in spatiotemporal instability of the output beam profile, if the flow is in the transition region between laminar and turbulent. For a variety of different geometries ranging from $L = 700d$ to $L = 10000d$, and for different gas species, we found that the peak of our (instability-) measurement variable is around a mean Reynolds number of 2400. Interestingly, the measured RIN drops for flow conditions in the fully turbulent regime at $Re>3500$ (Fig. 6(e)–6(i)). This could be associated with a smaller impact of the flow on the measured output beam stability due to averaging of many turbulences along the fiber and/or a change of the dimensions of the turbulences relative to the mode size and the observed cross-sectional area.

At this point, it is worth mentioning that the experimental conditions for the measurements presented in Fig. 6(e) are almost identical to those presented in Ref. [22]. This work describes a nonlinear pulse compression experiment, using ∼0.5 mJ ultrashort pulses at 2 µm wavelength. Herein, we have confirmed the observations reported by the authors (instabilities of the optical output at a pressure gradient >3 bar) and linked the onset in output beam profile fluctuations to the Reynolds number. The fact that we could confirm the instability threshold for high-intensity pulses shows the applicability of our measurement approach to explore the challenges to stable light propagation through differentially pumped capillaries. It is in this case (Fig. 6(e)) very clear to see how the instability suddenly increases for $Re>2040$, which is the cut-on for the transition region.

While every measurement presented in Fig. 6 shows a sudden increase in RIN for $Re>2040$, there are some individual differences. We believe that these differences are associated with the fact that we did not use any normalized, external triggers for inducing turbulences to the flow intentionally. This means that the emergence of puffs in the transition regime is highly dependent on random disturbances, e.g. originating from mechanical vibrations or variations in the flow from the gas supply, that are not controlled in the individual measurements.

## 4. Conclusion and outlook

In this work, we have observed that high throughputs can cause spatiotemporal instabilities in the output beam profile from differentially pumped capillaries. Resulting from numerical modelling and the experimental investigations presented herein, it is found that the onset of those strong instabilities is related to the Reynolds number $Re = 2040$, which indicates the beginning of the transition region between laminar and turbulent flow. Additionally, we found that the instabilities are most severe around $Re = 2400$. These results are purely observational, and while this work suggests a possible interpretation, the exact mechanism behind the beam profile fluctuations is to be further investigated.

Based on the observed correlation with the Reynolds number, we formulate guidelines to predict the onset of the spatiotemporal beam profile instabilities. For operation without turbulence, it is advisable to choose the upstream pressure $p_1$, the downstream pressure $p_0$, the fiber dimensions and the gas species in a way, such that the Reynolds number (Eq. (3)) does not exceed a value of 2000. The necessary throughput calculations can be performed using Eq. (1), since laminar flow is desired. This gives, for the special case that the pressure at the optical input of the capillary is negligible as compared to the upstream pressure $p_1$:

$$p_1 < 8 \cdot \left(\frac{RT\eta^2}{M_{\mathrm{mol}}}\right)^{\frac{1}{2}} \cdot \left(\frac{L}{d^3}\right)^{\frac{1}{2}} \cdot \sqrt{2000}. \qquad (4)$$

For the experimental conditions of Fig. 6(c), this gives $p_1<3.75$ bar, which agrees well with the experimental and numerical analysis. It is easy to see from Eq. (4) that for a nonlinear pulse compression system operating at the beginning of the transition region, a reduction in core size or increase in fiber length could suffice to reduce the throughput such that laminar flow is ensured. However, this comes at the cost of increased intensity and increased transmission losses, which is not desired either. Hence, such analytic formulas represent another important building block for the challenging design of high-energy, differentially pumped hollow fiber compressors and the formulas discussed herein are also applicable to a nonzero pressure at the optical input [13]. Furthermore, it is worth noting that the iterative numerical flow calculations allow obtaining the downstream pressure for chocked flow conditions, which is not necessarily equal to the ambient pressure in the evacuated recipient. Hence, these methods can be used to check the desired "zero" pressure condition at the optical input of a differentially pumped capillary.

In addition to the discussed implications for direct laser applications, we believe that the analysis of the optical output properties can be used as a novel tool for the investigation of volume flow through capillaries. By analyzing the light that is guided within the gas-filled hollow core fiber, we have not only confirmed the commonly known boundary between laminar and turbulent flow, but we have also characterized the frequency spectrum of the turbulences for our experimental case. Provided that our interpretations of this observational work can be confirmed by further studies, such measurements could be an interesting complement to the investigation of gas flow through capillaries in the context of mass spectrometry [26]. The methods discussed herein can be readily expanded to investigate the influence of externally triggered disturbances, to include the flow of liquids through hollow fibers [32] and to probe a hollow-core fiber perpendicular to the flow direction [33]. Because of the fast light propagation, optical probing of turbulences could be especially interesting for pump-probe measurements.

**Funding.** Fraunhofer-Gesellschaft (CAPS); H2020 European Research Council (835306).

**Acknowledgements.** The authors would like to thank Dr. César Jáuregui and Dr. Walter Wißdorf for fruitful discussions. This work was supported by the European Research Council (ERC) under the European Union's Horizon 2020 research and innovation program (grant 835306, SALT) and the Fraunhofer Cluster of Excellence Advanced Photon Sources (CAPS).

**Disclosures.** The authors declare no conflict of interest.

## References

1. F. Köttig, F. Tani, C. M. Biersach, J. C. Travers, and P. S. J. Russell, "Generation of microjoule pulses in the deep ultraviolet at megahertz repetition rates," Optica **4**(10), 1272 (2017).
2. J. C. Travers, T. F. Grigorova, C. Brahms, and F. Belli, "High-energy pulse self-compression and ultraviolet generation through soliton dynamics in hollow capillary fibres," Nat. Photonics **13**(8), 547–554 (2019).
3. D. Popmintchev, B. R. Galloway, M.-C. Chen, F. Dollar, C. A. Mancuso, A. Hankla, L. Miaja-Avila, G. O'Neil, J. M. Shaw, G. Fan, S. Ališauskas, G. Andriukaitis, T. Balčiunas, O. D. Mücke, A. Pugzlys, A. Baltuška, H. C. Kapteyn, T. Popmintchev, and M. M. Murnane, "Near- and Extended-Edge X-Ray-Absorption Fine-Structure Spectroscopy Using Ultrafast Coherent High-Order Harmonic Supercontinua," Phys. Rev. Lett. **120**(9), 093002 (2018).
4. A. Knebl, D. Yan, J. Popp, and T. Frosch, "Fiber enhanced Raman gas spectroscopy," TrAC Trends Anal. Chem. **103**, 230–238 (2018).
5. M. Nisoli, S. De Silvestri, and O. Svelto, "Generation of high energy 10 fs pulses by a new pulse compression technique," Appl. Phys. Lett. **68**(20), 2793–2795 (1996).
6. I. Matsushima, H. Yashiro, and T. Tomie, "10 kHz 40 W Ti:sapphire regenerative ring amplifier," Opt. Lett. **31**(13), 2066 (2006).
7. T. Nubbemeyer, M. Kaumanns, M. Ueffing, M. Gorjan, A. Alismail, H. Fattahi, J. Brons, O. Pronin, H. G. Barros, Z. Major, T. Metzger, D. Sutter, and F. Krausz, "1 kW, 200 mJ picosecond thin-disk laser system," Opt. Lett. **42**(7), 1381 (2017).
8. B. E. Schmidt, A. Hage, T. Mans, F. Légaré, and H. J. Wörner, "Highly stable, 54 mJ Yb-InnoSlab laser platform at 0.5 kW average power," Opt. Express **25**(15), 17549 (2017).
9. P. Russbueldt, T. Mans, J. Weitenberg, H. D. Hoffmann, and R. Poprawe, "Compact diode-pumped 1.1 kW Yb:YAG Innoslab femtosecond amplifier," Opt. Lett. **35**(24), 4169 (2010).
10. H. Stark, J. Buldt, M. Müller, A. Klenke, A. Tünnermann, and J. Limpert, "23 mJ high-power fiber CPA system using electro-optically controlled divided-pulse amplification," Opt. Lett. **44**(22), 5529 (2019).
11. X. Chen, A. Jullien, A. Malvache, L. Canova, A. Borot, A. Trisorio, C. G. Durfee, and R. Lopez-Martens, "Generation of 4.3 fs, 1 mJ laser pulses via compression of circularly polarized pulses in a gas-filled hollow-core fiber," Opt. Lett. **34**(10), 1588 (2009).

12. S. Bohman, A. Suda, T. Kanai, S. Yamaguchi, and K. Midorikawa, "Generation of 5.0 fs, 5.0 mJ pulses at 1 kHz using hollow-fiber pulse compression," Opt. Lett. **35**(11), 1887 (2010).
13. T. Nagy, S. Hädrich, P. Simon, A. Blumenstein, N. Walther, R. Klas, J. Buldt, H. Stark, S. Breitkopf, P. Jójárt, I. Seres, Z. Várallyay, T. Eidam, and J. Limpert, "Generation of three-cycle multi-millijoule laser pulses at 318 W average power," Optica **6**(11), 1423 (2019).
14. T. Nagy, M. Kretschmar, M. J. J. Vrakking, and A. Rouzée, "Generation of above-terawatt 1.5-cycle visible pulses at 1 kHz by post-compression in a hollow fiber," Opt. Lett. **45**(12), 3313 (2020).
15. A. Malvache, X. Chen, C. G. Durfee, A. Jullien, and R. Lopez-Martens, "Multi-mJ pulse compression in hollow fibers using circular polarization," Appl. Phys. B **104**(1), 5–9 (2011).
16. V. Cardin, N. Thiré, S. Beaulieu, V. Wanie, F. Légaré, and B. E. Schmidt, "0.42 TW 2-cycle pulses at 1.8 µm via hollow-core fiber compression," Appl. Phys. Lett. **107**(18), 181101 (2015).
17. G. Fan, T. Balčiūnas, T. Kanai, T. Flöry, G. Andriukaitis, B. E. Schmidt, F. Légaré, and A. Baltuška, "Hollow-core-waveguide compression of multi-millijoule CEP-stable 3.2 µm pulses," Optica **3**(12), 1308 (2016).
18. S. Hädrich, M. Kienel, M. Müller, A. Klenke, J. Rothhardt, R. Klas, T. Gottschall, T. Eidam, A. Drozdy, P. Jójárt, Z. Várallyay, E. Cormier, K. Osvay, A. Tünnermann, and J. Limpert, "Energetic sub-2-cycle laser with 216 W average power," Opt. Lett. **41**(18), 4332 (2016).
19. A. Suda, M. Hatayama, K. Nagasaka, and K. Midorikawa, "Generation of sub-10-fs, 5-mJ-optical pulses using a hollow fiber with a pressure gradient," Appl. Phys. Lett. **86**(11), 111116 (2005).
20. C. Vozzi, M. Nisoli, G. Sansone, S. Stagira, and S. De Silvestri, "Optimal spectral broadening in hollow-fiber compressor systems," Appl. Phys. B **80**(3), 285–289 (2005).
21. L. Lavenu, M. Natile, F. Guichard, X. Délen, M. Hanna, Y. Zaouter, and P. Georges, "High-power two-cycle ultrafast source based on hybrid nonlinear compression," Opt. Express **27**(3), 1958 (2019).
22. M. Gebhardt, C. Gaida, F. Stutzki, S. Hädrich, C. Jauregui, J. Limpert, and A. Tünnermann, "High average power nonlinear compression to 4 GW, sub-50 fs pulses at 2 µm wavelength," Opt. Lett. **42**(4), 747 (2017).
23. G. Tempea and T. Brabec, "Theory of self-focusing in a hollow waveguide," Opt. Lett. **23**(10), 762 (1998).
24. K. Avila, D. Moxey, A. de Lozar, M. Avila, D. Barkley, and B. Hof, "The Onset of Turbulence in Pipe Flow," Science **333**(6039), 192–196 (2011).
25. B. Eckhardt, "A Critical Point for Turbulence," Science **333**(6039), 165–166 (2011).
26. W. Wißdorf, D. Müller, Y. Brachthäuser, M. Langner, V. Derpmann, S. Klopotowski, C. Polaczek, H. Kersten, K. Brockmann, and T. Benter, "Gas Flow Dynamics in Inlet Capillaries: Evidence for non Laminar Conditions," J. Am. Soc. Mass Spectrom. **27**(9), 1550–1563 (2016).
27. H.-J. Otto, F. Stutzki, F. Jansen, T. Eidam, C. Jauregui, J. Limpert, and A. Tünnermann, "Temporal dynamics of mode instabilities in high-power fiber lasers and amplifiers," Opt. Express **20**(14), 15710 (2012).
28. R. G. Livesey, "Solution methods for gas flow in ducts through the whole pressure regime," Vacuum **76**(1), 101–107 (2004).
29. M. Wutz, H. Adam, and W. Walcher, *Theorie Und Praxis Der Vakuumtechnik* (Vieweg + Teubner Verlag, 1992), **224O**(1).
30. W. Wutz, M. Adam, and H. Walcher, *Wutz Handbuch Vakuumtechnik* (Vieweg + Teubner Verlag, 2004).
31. I. J. Wygnanski and F. H. Champagne, "On transition in a pipe. Part 1. The origin of puffs and slugs and the flow in a turbulent slug," J. Fluid Mech. **59**(2), 281–335 (1973).
32. M. Chemnitz, M. Gebhardt, C. Gaida, F. Stutzki, J. Kobelke, J. Limpert, A. Tünnermann, and M. A. Schmidt, "Hybrid soliton dynamics in liquid-core fibres," Nat. Commun. **8**(1), 42 (2017).
33. J. R. Koehler, F. Köttig, B. M. Trabold, F. Tani, and P. S. J. Russell, "Long-Lived Refractive-Index Changes Induced by Femtosecond Ionization in Gas-Filled Single-Ring Photonic-Crystal Fibers," Phys. Rev. Appl. **10**(6), 064020 (2018).